\documentclass[cite,doublecol,figures]{epl2}
\usepackage{epsfig,color}
\usepackage{amsfonts}
\usepackage{amssymb}
\usepackage{amsmath}
\usepackage{verbatim}
\definecolor{brblue}{rgb}{0,1,1}
\definecolor{orange}{rgb}{1,0.5,0}
\usepackage[english]{babel}

\newcommand{\Livcu}{LiVCuO$_4$}

\newcommand{\Hubbard}{\textsc{Hubbard}}

\newcommand{\IFW}{Leibniz-Institut f{\"u}r Festk\"orper- und Werkstoffforschung (IFW) Dresden, D-01171 Dresden, PF 270116, Germany\\}
\newcommand{\ITPM}{Institut f\"ur Theoretische Physik, Universit\"at Magdeburg,  Magdeburg, Germany\\}

\newcommand{\IMSKiev}{Institute for Problems of Materials Science Krzhizhanovskogo 3, 03180 Kiev, Ukraine\\}
\newcommand{\MPICPfS}{Max-Planck-Institut f\"ur Chemische Physik fester Stoffe, Dresden, Germany\\}
\newcommand{\Prague}{Institute of  Physics, ASCR, Prague, Czech Republic\\}

\title{The strength of 
frustration and quantum fluctuations in  \Livcu }

\shorttitle{Frustration and \Livcu }

\author{Satoshi Nishimoto\inst{1}, 
Stefan-Ludwig Drechsler\inst{1}\thanks{Corresponding author, E-mail: \email{s.l.drechsler@ifw-dresden.de}}, 
Roman Kuzian\inst{1,2},
Johannes Richter\inst{3},\\
Ji\v{r}i M\'alek\inst{1,5},
Miriam Schmitt\inst{4},
Jeroen van den Brink \inst{1}, {\lowercase{and}}  Helge Rosner\inst{4}  
\shortauthor{\sc
Satoshi Nishimoto \etal }}

\institute{ \inst{1}\IFW \inst{2}\IMSKiev \inst{3}\ITPM 
\inst{4}\MPICPfS
\inst{5}\Prague

}


\abstract{We present an empirical
and microscopical
analysis of the 
main in-chain
exchange constants
of the edge-shared frustratrated chain cuprate
\Livcu = LiCuVO$_4$ \ 
with a ferromagnetic nearest neighbour coupling
which clearly exceeds the antiferromagnetic (AFM) next-nearest neighbour 
exchange $J_2$.
The measured saturation field is significantly affected by a weak  3D AFM interchain
coupling leaving room for a possible Bose-Einstein condensation for several T below.
 The obtained
exchange parameters are in agreement with the results for a realistic five-band
extended \Hubbard{} Cu $3d$ O $2p$ model, LSDA+$U$ predictions as well as with inelastic
neutron and magnetization data. The single chain frustration rate
$\alpha=J_2/|J_1|\approx 0.75$, including all error bars, is definitely smaller than 1.0
which correspond to strongly coupled interpenetrating AFM Heisenberg chains
in contrast with opposite statements in the literature. 
A proper account of strong quantum fluctuations and frustration
is necessary for a correct assignment of the exchange integrals.
which cannot be achieved by a simple renormalization of $J_2$ from 
spin-wave theory.}

\begin{document}

\maketitle

\section{{\bf 1. INTRODUCTION\/}}
\Livcu $\equiv$ LiCuVO$_4$ is one of the first \cite{Hoppe1970,Gibson04} and 
rather  frequently studied spin-chain 
compounds among
edge-shared cuprates 
\begin{figure}[b!]
\includegraphics[width=0.40\textwidth]{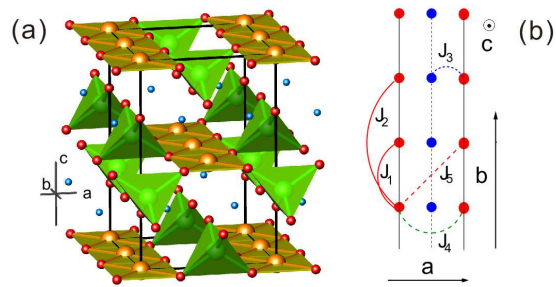}
\caption{(Color) (a):
the 
crystallographic structure of \Livcu \ comprises two AFM coupled
CuO$_2$ spin-chains per unit cell running  along the $b$-axis 
(orange {\large \color{orange} $\bullet$} -- Cu$^{2+}$, 
 red    {\large \color{red}    $\bullet$} -- O$^{2-}$, 
bright blue {\large \color{brblue} $\bullet$}
 -- Li$^{+}$).  
(b):
the main
 in-- and inter-chain exchange 
paths, $J_1$, $J_2$, and $J_3$,$J_4$, $J_5$
marked by solid
red arcs,
broken red line,
blue and green
arcs , respectively.  
} 
\label{fig::Struct}
\end{figure}
\cite{Enderle05,Enderle10,Buttgen10,Enderle11,Drechsler2007JP}. 
Recently it became especially interesting due to the observation of 
multiferroicity \cite{Yasui08,Moskvin08,Sato10} 
and due to a possible realization of quantum spin nematics and related 
Bose-Einstein condensation
of two-magnon bound states
in high magnetic fields 
\cite{Zhitomirsky10,Svistov11,Drechsler2007JMMM}.
Both phenomena
are still poorly understood and a precise knowledge of the main exchange
interactions is of key importance
to attack such complex problems in a realistic way.
Unfortunately, there is no consensus about the magnitude of these
couplings and in particular on the value of the in-chain 
frustration parameter 
$\alpha=J_2/|J_1|$, with $J_1<0$ as the ferromagnetic (FM)
 nearest neighbor (NN) 
  and $J_2$ the 
antiferromagnetic (AFM) next-nearest neighbor (NNN)-coupling in chain 
direction $b$ (see Fig.\ 1).
So far, in various studies
 $0.5 \leq \alpha \leq 2.2$ 
and even  
above 5.5
have been predicted/reported.\cite{Sirker10,Drechsler11,Enderle10,Enderle05,Koo11}
Keeping in mind the weak interchain coupling,
\begin{figure*}[t]
\begin{center}
\includegraphics[width=0.85\textwidth]{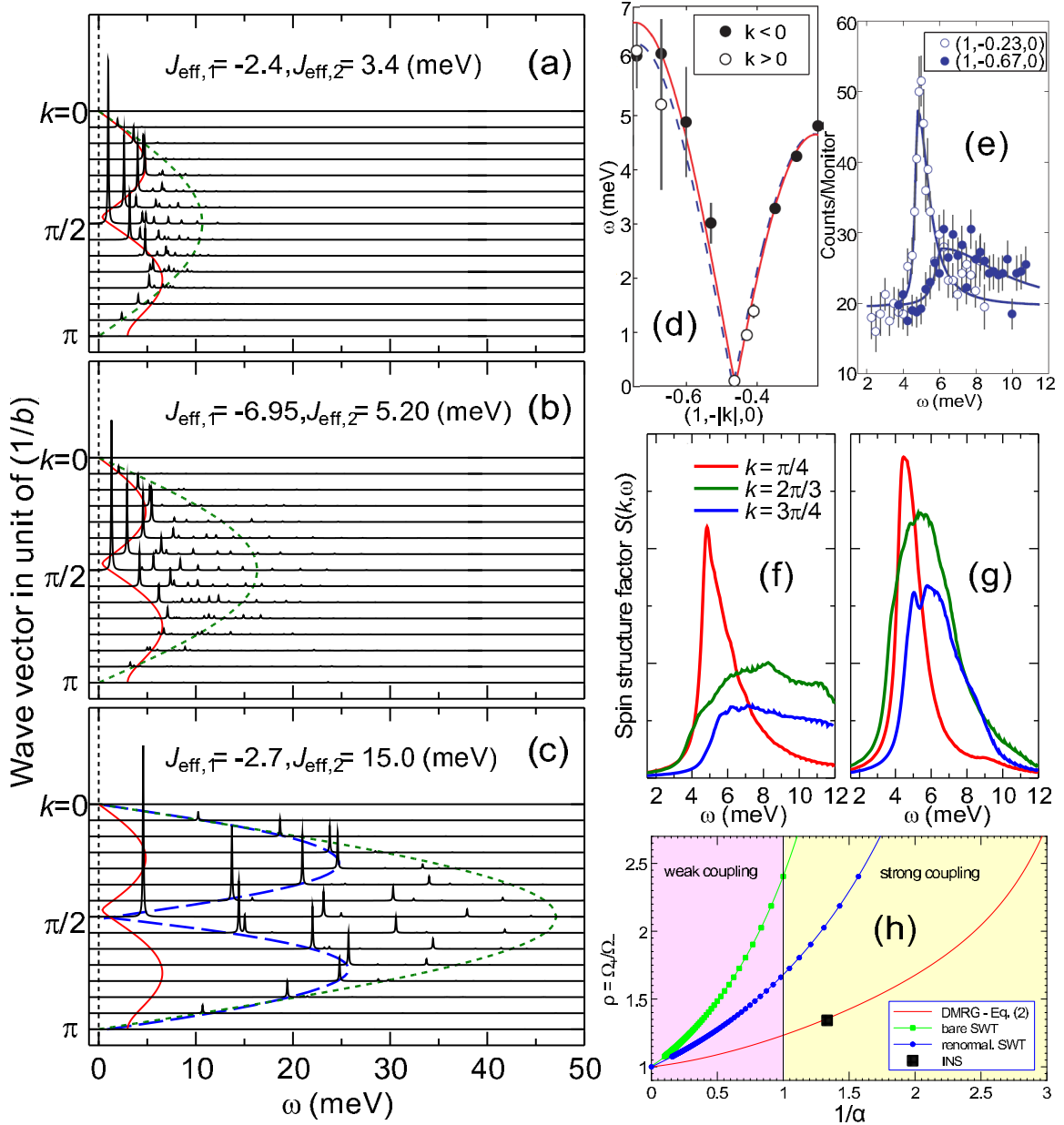}
\caption{(Color) Left: 
 Magnetic dynamical structure factor $S(k,\omega)$ 
from exact diagonalizations for periodic
chains 
and $L=28$ sites
for the parameter sets proposed in Ref.\  \cite{Enderle10} (a), 
our recent findings \cite{Drechsler11}  (b) and that
by Koo {\it et al.} 
\cite{Koo11} for $U=5$~eV (see Tab.\ 1)
(c). Right: experimental {\it asymmetric}
dispersion along the chain direction in between $\Omega_+$
and $\Omega_-$ (see Fig.\ 2(h) and
Eq.\ (2))
and INS-intensities from Ref.\ \cite{Enderle10} (d,e), 
calculated $S(k,\omega)$ normalized to the 
static structure factor $S(k)$
(i.e.\ the
$\omega$-integrated dynamical one)
for a single chain with our parameters
(f) and that of Ref.\  \cite{Enderle10} 
(g), 
and the ratio of the peak
positions at the first and second 
maximum of $S(k,\omega)$ 
for the transferred momenta near $k$=1/4 an 3/4, respectively, 
(h)
approximately given by the red curves in (a-c).
In all DMRG-
calculations
for chains with open boundary conditions, 
$L=96$ sites, and a Lorentzian broadening at half-width
$\Gamma=0.3$~meV
 have
 been employed.
}
\end{center}
\label{fig::INS}
\end{figure*}
\noindent
the single-chain can be viewed also as two
interacting and interpenetrating simple AFM Heisenberg chains (AHC) or 
equivalently
as a single
zigzag ladder. Then, one is left with a weak ($\alpha >1$)
or 
a strong coupling  
scenario 
in the opposite case $0< \alpha <1$
 because $1/\alpha$ provides  a direct
measure 
of the FM "interchain"-coupling within a zigzag-ladder or between interpenetrating 
AHC.
Here, we give a more comprehensive
presentation 
of arguments {\it against} weak coupling scenarios 
than 
in our 
short
Comment \cite{Drechsler11} on
Ref.\ \cite{Enderle10} and rebut in more detail also
arguments put forward in Ref.\ 
\cite{Enderle11}. In addition,
the striking discrepancy of a very
recent parameter set obtained in Ref.\ \cite{Koo11}  with 
available experimental data, including also 
Refs.\ \cite{Enderle10,Enderle11}
will be shown, too. Finally, the reasons of the incorrect
parameter assignment given in Refs.\  \cite{Enderle05,Enderle10,Koo11}
 are explained in terms of an inproper handling of strong
 quantum fluctuations (SQF). Clear evidence for SQF
  comes from the small values of the ordered
 magnetic moment of 0.31$\mu_{\rm B}$ and the low
 Ne\'el-temperature
$T_{\rm N}\approx 2.4$~K \cite{Gibson04,Sato10}.

 \section{\bf 2.\  PHENOMENOLOGICAL ANALYSIS}
The low-energy inelastic neutron scattering (INS) data
\cite{Enderle05} 
have been fitted in terms
of simple (bare) 3D
spin wave theory (BSWT), i.e.\
{\it without} any renormalization or account of quantum fluctuations
(see Figs.\ 2 (a-c,d,e)). As a result one arrives
at $J_2^{\rm BSWT} \approx$~5.6 meV and $J_1^{\rm BSWT} \approx-1.6$~meV 
despite weak interchain couplings among them $J_5^{BSWT}=-0.4$~meV
has been claimed to be the predominant interaction responsible for the 
in-phase ordering of spirals in the magnetically ordered state below 
$T_{\rm N}$ (see Tab.\ 1).
Then based on these and recent
high-energy data 
(20~meV$>
\omega >
0.5$~meV data)\cite{Enderle10} analyzed by
means of a random phase approximation (RPA)
approach for the account of the coupling between the AHC,
an effective 1D model has been proposed in Ref.\ \cite{Enderle10}:
\begin{equation}
J_{\rm eff, 2}\approx (2/\pi)J^{\rm SWT}_2\quad \mbox{and} \quad
J_{\rm eff, 1}
\approx J_1^{\rm SWT}+2J_5^{\rm SWT}, 
\end{equation}
Thereby quantum fluctuations 
have been taken into account by the prefactor $2/\pi$
as for free AHC
in accord with the implicite {\it a priori}
assumption of a weak coupling between the AHC,
in other words
any renormalization related
 to the coupling
 $J_{\rm eff,1}$
between the AHC
has been almost 
ignored.
Such an assumption seems to be to not justified from a general
many-body theory point of 
view.
However, following our recent work
\cite{Drechsler11} we will show here in more 
detail that the  in-chain 
exchange integrals $J_1$ and $J_2$, respectively,
are significantly different from those suggested
in Refs.\cite{Enderle05,Enderle10,Enderle11,Koo11}.
\begin{table}[t]
\caption{Exchange integrals (in meV) as extracted from INS data using 
bare spin
wave theory (BSWT), 
ad hoc renormalized spin wave theory (RSWT)\cite{Enderle05} 
or RPA \cite{Enderle10} as compared with parameters derived from
mapping of
microscopic models (see sect.\ 3). In all 
LDA(GGA)+$U$ shown 
calculations
a value $U-J=5$~eV
appropriate 
for edge-shared cuprates   
has been used.  }
\label{tab::ExchangeParameters}
\begin{center}
\begin{tabular}{r|c|c|c|c|c}
\hline
\hline
& $J_1$        & $J_2$         & $\alpha$ & $J_4$   & $J_5$ \\
 \hline
BSWT \cite{Enderle05}               & $-$1.6 & 5.59   & 3.49  
        & 0.01 
& $-$0.4  \\
RSWT \cite{Enderle05} & $-$1.6 & 3.56 & 2.23
 &&\\
RPA\cite{Enderle10}  & $-$2.4& 3.4 &1.42 & &   \\
\hline
present work    & $-$6.95       & 5.2  & 0.75           &         &              \\
3$d$O2$p$ optics     & $-$6.31      & 5.05             &      0.8       &  &         \\
GGA+$U$\cite{Koo11}  & $-$2.7       & 15.0  &     5.55  &   -1.31   & 0.16           \\
LSDA+$U$ & $-$8.5    & 7.05                 &   0.82  & &  \\
GGA+$U$  & -6.4     &  5.45 &
 0.85               &  &  \\
\hline
\hline
\end{tabular}
\end{center}
\end{table}
Furthermore it has been claimed
that the FM coupling $J_1$ can
  be fixed
at its bare value $J^{\rm BSWT}_1=J^{\rm RSWT}_1$. 
This would yield $J^{\rm RSWT}_2\approx 3.57$~meV close to the result
of a phemonelogical
 RPA-based
 description of the problem: $J^{\rm RPA}_{2}\approx $3.4~meV \cite{Enderle10}.
In passing through we note that the value 
predicted by Koo {\it et al.}
\cite{Koo11} 
exceeds that value very much
by a factor larger than three.
If the interchain coupling is of less relevance for the "high-energy"
physics, the claimed 2$J_5\approx -0.8$~meV should be {\it added}
to $J^{\rm RPA}_1=-2.4$~meV only for low-energy problems such as
thermodynamics, i.e.\ 
relevant for the saturation field and the magnetization
or the determination of the spiral's pitch angle. 
With such a more convincing empirical RPA affected renormalization one 
would already 
arrive at $\alpha =1.063$ close to the strong coupling boarder line.
Up to now all considerations were based on the assumption that the 
FM $J_1$ remains fixed. However, 
field-theory flow-equations based
approaches \cite{Nersesyan99}
valid at $\alpha \gg 1$ point 
to strong
coupling renormalizations. As a
consequence, $J_1$ might change considerably and $\alpha$ is further 
scaled down. In fact, such a tendency would be compatible with our DMRG
\cite{White92} 
results 
(see also
below and Tab.\ 1) : $J_{\rm eff,1}=-6.95$~meV, $J_{\rm eff, 2}= 5.2$~eV,
and $\alpha \approx 0.75$ \cite{Drechsler11}. 
If one adopts the BSWT-parameters as a reasonable starting point,
our results should be interpreted as a strong {\it upward} renormalization
of both $|J_1|$ and a moderate for $J_2$, too,

Turning first to the low-energy INS-data \cite{Enderle05}, 
we start with the two extrema  
of a one magnon excitation $\Omega_{-}$ and $\Omega_+$, i.e.\
the peak positions near the transferred momenta 
 $k=1/4$
and $k=3/4$.
This is the lowest  two-spinon excitation (2SE) reproduced approximately
by the BWST-fit taken from Ref.\ \cite{Enderle05,Enderle10}.
Its dispersion is sketched 
by the red curves
in Fig.\ 2 (a-c). Although the 
maximum corresponding to $\Omega_+$ is broad, 
the asymmetry with respect to $\Omega_-$ 
quantified by the dynamical
asymmetry parameter
$\rho=\Omega_+/\Omega_- \geq 1$ 
is clearly visible in the 
INS-data 
in contrast
with the set proposed in Ref.\ \cite{Koo11} where 
$\rho \approx 1$
would occur. On  absolute scale a discrepancy
by factor exceeding three between the experimental and the predicted 
dispersion 
is observed (see 
 Fig\ 2(c) which can be traced back
to artificially large values of $J_2$
(see Tab.\ 1). 

From the experiment,
 Fig.\ 2 (d),
   one reads off $\Omega_{-}=4.84$~meV. Taking 
$\Omega_+=6.4$~meV
one estimates $\rho \approx 1.32$.
$\rho$
can be obtained from fitting our dynamical 
DMRG \cite{Jeckelmann02}
results for 
$0.3 \leq \alpha \leq 3$ and 
long chains
with $L=96$ sites
\begin{eqnarray}
 \frac{\Omega_+}{J_2}&=&
\frac{\pi}{2}+0.0338x-0.302x^2+0.0831x^3-0.00699x^4, 
\nonumber
\\
\frac{\Omega_-}{J_2}&=&\frac{\pi}{2}-0.143x-0.534x^2+0.279x^3-0.0589x^4 +
 \nonumber
 \\
 &&+0.00465x^5, 
\label{INS}
\end{eqnarray}
where the coupling strength
$x=1/\alpha=|J_1|/J_2$ has been introduced.
The relation $\alpha=f(\rho)$ 
provides a convenient highly 
sensitive measure of
the interaction regime which is heavily
affected by the strong  quantum fluctuations.
The function $\alpha(\rho)$ 
is depicted in 
Fig.\ 2 (h). One realizes 
excellent agreement 
with  $\alpha=0.75$
derived in our previous paper where instead 
$\Omega_{-}$ and the 
relative
magnetization
curve $M(H)/M_{\rm s}$ as a function of $H/H_{\rm s}$ 
at low temperature have been 
employed \cite{Drechsler11}.
Notice the large deviations if the BSWT or the RSWT would be applied
to extract $\alpha$. Taking $\Omega^{BSWT}_+\approx 6.7$~meV 
and $\Omega^{BSWT}_-\approx 4.75$~meV 
from 
Figs.\ 2 (d,e) which yields $\alpha^{\rm BSWT}\approx 2.3$
almost 
consistent with 2.2 stated in Ref.\ \cite{Enderle05}. Using the RPA-derived
values one arrives at $\alpha^{\rm RPA}$
about 1.42 again in formal consistency with \cite{Enderle10}.
The strong deviations of both values from our DMRG-based value
clearly show the inapplicability of simple spin-wave theory based estimates.
The physical reason is the incorrect treatment of strong quantum fluctations
in the title compound which manifest themselves also 
in a small magnetic moment as mentioned above and in relatively 
large pitch angles (see below).

Finally, considering briefly the calculated and the
experimental INS
intensities, at present 
only few comparisons are possible due to lacking 
publication of 
experimental spectra.
Nevertheless,
 comparing e.g.\    
 the available data shown in Figs.\ 2 (d,e) 
one realizes that our set provides a better description of the 
intensity at large transferred momenta (Fig. 2(b)) as compared with that 
of Ref.\ \cite{Enderle10}.
A comparison of the 
theoretical
shapes
with more INS
spectra
would be  helpful to reduce our error bars. 


If one adopts that the experimental magnetization data up to the so-called 
field $H_{c3}\approx 40.5 \pm 0.2$~T ($H \parallel c) $
where
 the peak in 
dM/dH occurs \cite{Svistov11} is well-described by an effective 1D model,
one arrives at the curves shown in Fig.\ 3.
Notice the strong deviation of the weak-coupling proposal
by Koo {\it et al.}\cite{Koo11}.
\begin{figure}[b!]
\begin{center}
\includegraphics[width=5.2cm,]{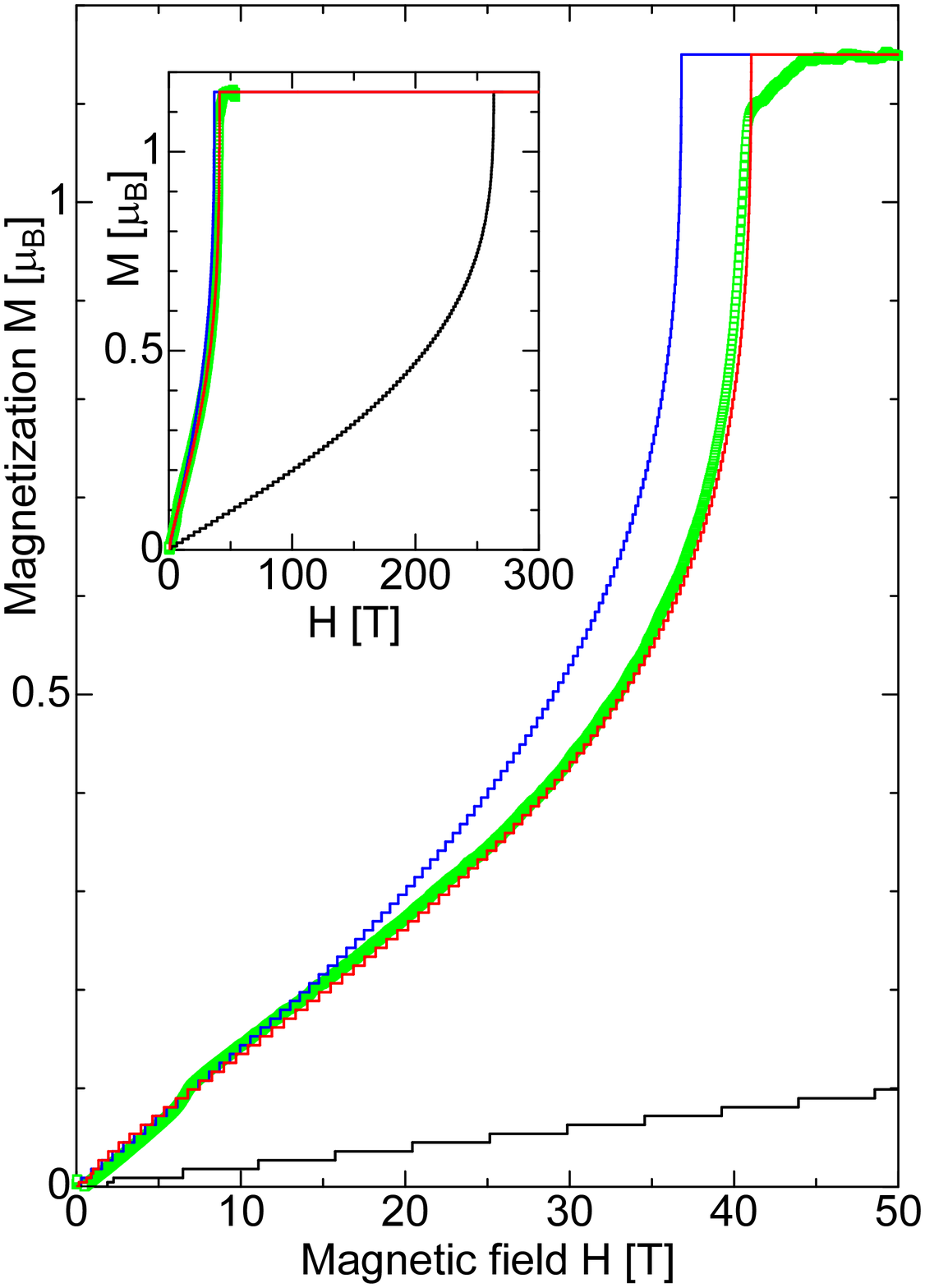}
\caption{(Color) Magnetization vs. applied magnetic field.
Experiment (\textcolor{green}{$\circ $} from
Ref.\ \cite{Svistov11}), theory: blue line - set of 
Ref.\ \cite{Enderle10}, red line -
our set \cite{Drechsler11} and black  line- the $U=5$~eV- set
from Ref.\ \cite{Koo11} and
$g=2.3$  for all sets. The
DMRG calculations were
 performed
for $L=512$ sites at $T=0$. Inset: 'entire' field range.
}
\label{f1} 
\end{center}
\end{figure}
Then, a dominant
 FM interchain coupling as proposed 
in Refs.\ \cite{Enderle10,Enderle05} cannot be reconcoiled with these 
experimental data since for such a coupling
the 3D saturation field is smaller than its 1D counterpart \cite{Kuzian07}.
Also the spin susceptibility $\chi(T)$ is within an RPA 
approach for the interchain coupling  best described by a total AFM 
interchain coupling. Anyhow, a detailed discussion of $\chi_{\rm s}(T)$
including also a consideration of the background susceptibility
$\chi_0$
will be given elsewhere.

\begin{figure}[b!]
\begin{center}
\includegraphics[width=6.0cm,]{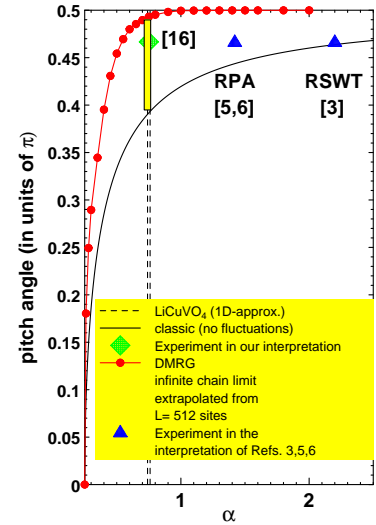}
\caption{(Color) Pitch angle for a single chain from the maximum of the 
static 
magnetic structure factor S(q) in comparison with experiment,
Refs.\ \cite{Enderle05, Enderle10,Enderle11},
and Eq.\ (4) (black line).}
\label{f1} 
\end{center}
\end{figure}
As mentioned above
the presence of strong quantum fluctuations is 
evidenced by the small magnetic momentum of
0.31$\mu_{\mbox{\tiny B}}$ and a 
low Ne\'el-temperature $T_{\rm N}=2.4$~K. Both values should be 
compared e.g.\ with three to four times large values for the
sister compound Li$_2$CuO$_2$ \cite{Chung03,Sapina90} caused by
a relatively strong interchain coupling \cite{Lorenz09}. In addition
its small $\alpha \approx 0.32$ is also helpful to suppress
SQF. As a consequence the spiral state 
is significantly 
driven towards almost decoupled AHC
the corresponding collinear Ne\'el state, of
each AHC
i.e.\  the
experimental pitch angle $\phi=84^\circ$ analyzed within the BSWT or
RSWT
results in strongly overestimated $\alpha$-values.
This is illustrated in Figs.\ 4 and 5, where the maximum of the static
magnetic structure factor
is depicted as a function of $\alpha $ for the cases of
a single frustrated $J_1$-$J_2$ 
chain and a coupled pair of them, respectively. Already
the latter is expected
to provide a
reasonable insight into
the real 
quasi-1D situation.
This point of view is supported  by 
a detailed
comparison with 
coupled cluster calculations
to be reported elsewhere.
Thus, for instance 
 in the case of a planar arrangement
of chains (i.\ e.\ a dominant 2D-interchain coupling as in the model 
adopted  in Refs.\ \cite{Enderle05,Enderle10,Zhitomirsky10, Svistov11})
the effective interchain interactions $J_5^*$  and $J_4^*$ correspond 
approximately to
\begin{equation}
J_5^*=2J_5; \qquad J_4^*=2J_4 \quad .
\end{equation}
\begin{figure}[t!]
\begin{center}
\includegraphics[width=7.0cm]{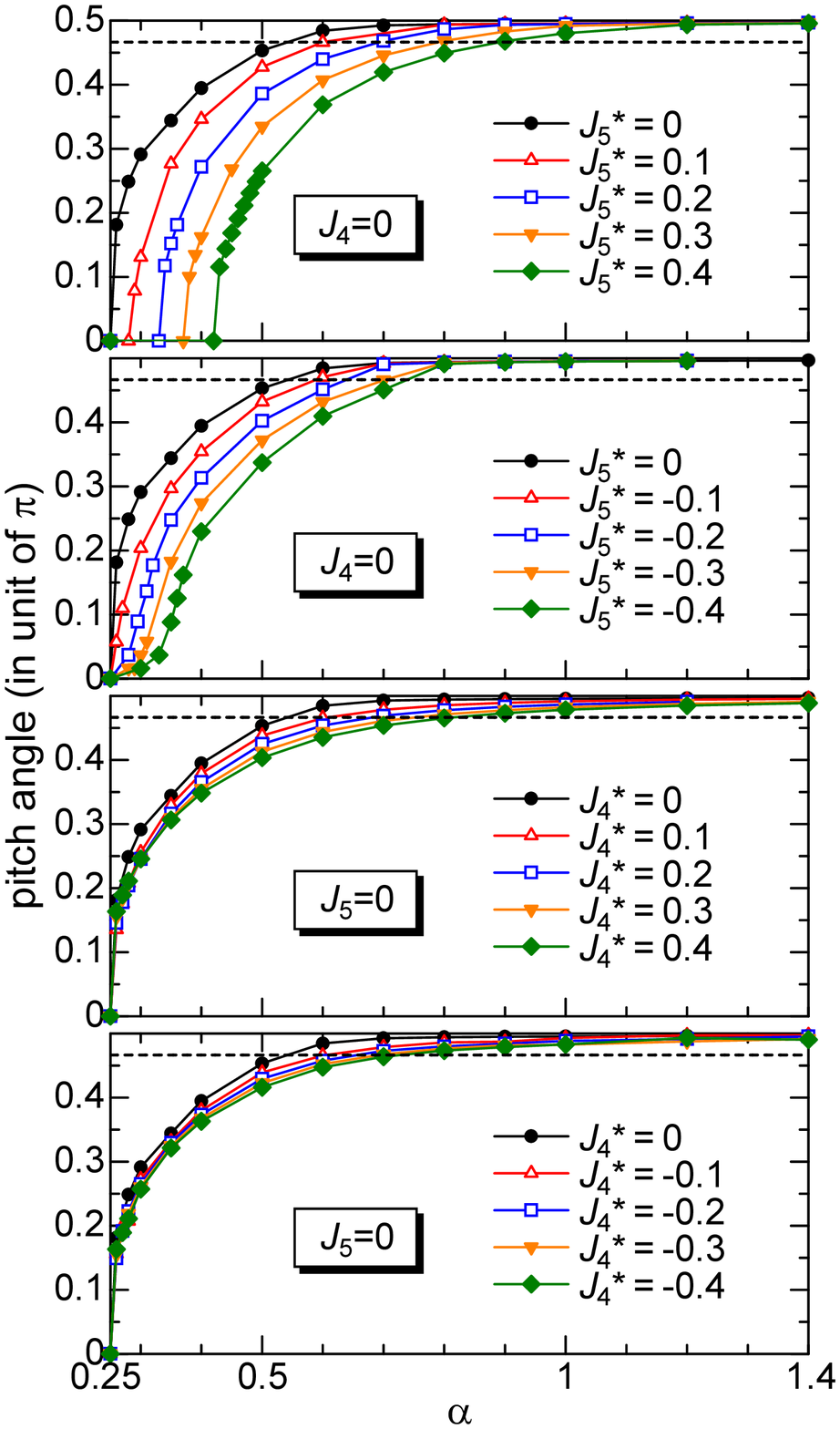}
\caption{(color) Pitch angle for two coupled chains with various
types of interchain coupling in units of $|J_1|$ 
(see Fig.\ 1(b)).
The corresponding
propagation vector 
has been estimated from the maximum of the calculated (DMRG)
static magnetic structure factor $S(q)$
 for $L=192$ sites
in total. The experimental value
amounts about 0.467  (dashed line) \cite{Gibson04}. 
}
\end{center}
\label{f5} 
\end{figure}
Notice the striking failure
of the classical curve 
especially
for large $\alpha$ (see Fig.\ 4).
Such an effect was first addressed in Ref.\ \cite{Bursill95} in the 1D-case
by means of DMRG and in Ref.\ \cite{Zinke09} for a plane of perpendicularly
coupled chains by means of a coupled cluster approach.
We stress that the experimental value of the pitch
\cite{Gibson04} is reproduced for $\alpha \leq 1$, only, independently 
of the details of the weak interchain coupling.
Adopting the in-chain
parameters and the leading interchain coupling $J_5\approx -0.4$~meV 
suggested in
Ref.\ \cite{Enderle05},
one estimates from Fig.\ 5 a pitch angle of 
89.58$^\circ$
($\alpha =2.22$)
and of 
89.43$^\circ$
 in the case of the RPA derived set ( $\alpha=1.42$)
in contrast to 84$^\circ$ known experimentally \cite{Gibson04}.
Thus, the measured $\phi$
points clearly to a strong coupling regime
in contrast to opposite statements 
in Refs.\
\cite{Enderle05,Enderle10,Koo11}.
Naturally, the SWT derived $J$'s obey nearly the classic
relation, only, 
 (i.e.\ completely 
 ignoring the strong quantum
fluctuations)
\begin{equation}
\phi=\cos^{-1}(0.25/\alpha) ,
\end{equation}
yielding 83.53$^\circ$ 
for $\alpha=2.22$ \cite{Enderle05} (the small deviations from 84$^\circ$ 
result from the weak interchain
coupling ignored in Eq.\ (4) for the sake of simplicity)
and 87.42$^\circ$ for $\alpha=5.55$ \cite{Koo11}.
What matters here is not the absolute value of $\phi$, but the difference
$\pi/2-\phi$ which differs by two orders of magnitude between the quantum 
and the classic case \cite{Aligia00}.
Thus, the attempt to describe the spin dynamics in a quasi-classical
way is the main reason for the inproper
 assignment  in  
Ref.\ \cite{Enderle05}.

\section{{\bf 3.\ MICROSCOPIC ANALYSIS}}
Turning to a microscopic analysis,
we compare
our (DMRG)
INS derived $J$'s
with those from 
two independent microscopical 
approaches:
(i) analyzing high-energy spectra from 
EELS, optical conductivity $\sigma(\omega)$ or RIXS data
within strongly correlated extended multiband Hubbard models
and a subsequent mapping of their spin-states onto
the corresponding states of
a spin-Hamiltonian, i.e.\ the 1D $J_1$-$J_2$ model under consideration.
The results are shown 
in Tab.\ 1 and Fig.\ 6. 
(ii) extracting these exchange parameters from
total energy calculations 
of various prepared artificially magnetically ordered states
(see e.g.\ \cite{Koo11}). 
\begin{figure}[t!]
\begin{center}
\includegraphics[width=0.42\textwidth]{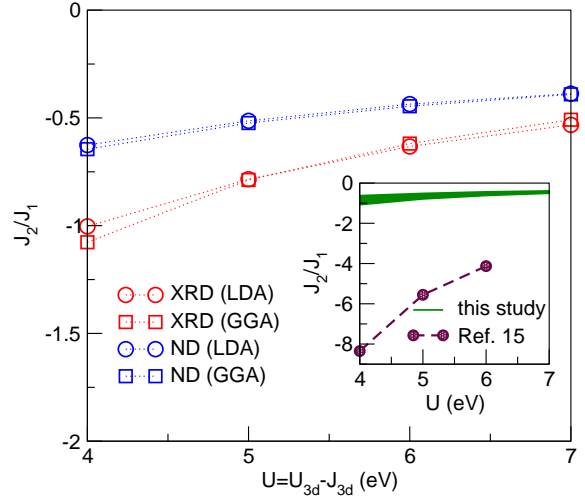}
\caption{(Color) 
Frustration parameter $\alpha$ vs.\ effective
on-site repulsion $U$ for two different crystal structure refinements 
from x-ray  (XRD) and neutron diffraction (ND). For each 
structure, two versions of the exchange correlation potential have 
been applied (LDA+$U$ and GGA+$U$) within the around mean field 
double counting scheme and using
$J_{3d}=1$~eV.
Inset:  The range of the obtained $\alpha$ values from our 
calculations in 
dependence of $U$ compared
with the results of Koo {\it et al.} \cite{Koo11}.}
\end{center}
\label{fig: 6}
\end{figure}
A mapping from a  Cu3$d$ O2$p$ five-band Hubbard model with usual parameters
which describes the $T$-dependent dielectric response
\cite{Malek08,Matiks09} onto a $J_1$-$J_2$ spin-1/2 model
yields a sizeable $J_1$=$ -$6.3~meV and $J_2$=5.05~meV.
We stress that in all closely related sister compounds \cite{Drechsler05}
with a
 Cu-O-Cu bond angle $\stackrel{<}{\sim}$95$^\circ$ sizeable 
FM $|J_1|$-values $\gg$ 1.6~meV have been found in fitting various data:
Li$_2$CuO$_2$: $J_1=-$1~9.6~meV (INS \cite{Lorenz09}), 
Ca$_2$Y$_2$Cu$_5$O$_{10}$: $J_1=-~14.7$~meV (INS \cite{Kuzian11}), 
Li$_2$ZrCuO$_4$: $J_1=-23.7$~meV ($\chi(T)$, $c_p$ \cite{Drechsler07PRL,Sirker10}).
In particular, also for \Livcu \ the $T$-dependent 
optical conductivityidata \cite{Matiks09}
obtained from ellipsometry measurements can be well
fitted  within a five-band
Cu 3$d$ O 2$p$ extended Hubbard model on chain-clusters with up to
six CuO$_4$-plaquettes connected by edge-sharing. Thereby 
$U_d=8$~eV, $U_p=4.1$~eV
 $K_{pd}=65$~meV, $\Delta_{pd}=3.82$~eV 
etc.\ has been used. As a result one arrives at 
in-chain $J$'s
close to the INS-derived ones:
$\alpha=0.8$ and $J_2$=5.1~meV (see Tab.\ 1). The value of $J_1$ is 
sensitive to the magnitude of the direct FM exchange $K_{pd}$
whereas $J_2$ is mainly sensitive to the in-chain O-O transfer integrals.
Thereby $| J_1| \propto  K_{pd}$ holds approximately.
 Notice that the contribution of $K_{pd}$ is much more
important for the large negative (FM) value of $J_1$ than that of the intra-atomic
FM Hund's rule coupling on O.
In the past $K_{pd}$ has been used mostly as a fitting
parameter 
ranging from 50 to 110~meV for CuGeO$_3$ \cite{Mizuno1998,Braden1996}.
A reliable
$J_1$-value derived from an
INS analysis as
reported here is helpful
to restrict its value 
and opens a door for systematic studies of this important 
exchange and useful comparisons with other sister 
compounds \cite{Drechsler05}. In fact, our empirical 
value corresponds to
130~meV for a $\sigma $-Cu-O bond to be compared with 180~meV
estimated for that case and a cuprate plane in high-$T_c$ superconductors
\cite{Hybertsen92}.

Considering
 the total energies of various magnetic states 
the main exchange integrals can be also extracted from LDA+$U$ or GGA+$U$
calculations. Thereby the results
depend mainly on a single parameter
$U=U_{3d}-J_{3d}$, where $J_{3d}\approx 1$~eV denotes the Hund's rule 
coupling that is rather precisely known for transition metals. For 
both approximations, we 
calculated the exchange integrals $J_1$ and $J_2$ and their ratio 
$\alpha$ for the two different crystal structures refined from x-ray 
diffraction  and neutron diffraction 
(labeled XRD and
NRD, respectively, in Fig.\ 6 and below). As the key parameter, the 
resulting $\alpha$ for different values of $U$ is  presented in Fig.\ 6 
and given 
in Table 1 ($U=5$ eV). The graph indicates only small differences for 
the two crystal structure solutions, but essentially no difference 
for the two choices of the exchange-correlation potential (LDA+$U$ vs. 
GGA+$U$).
For realistic parameters which describe successfully other edge-shared
chain cuprates one arrives again at $\alpha <1$ in sharp contrast to
Koo {\it et al.} \cite{Koo11} who obtained
 unusually large $J_2$-
and $\alpha $-values not compatible with 
 the
observed pitch angle \cite{Gibson04}, the
restricted
two-spinon continuum,
 and an obviously
{\it asymmetric}
INS spectrum
\cite{Enderle10}.
Presumably it is a consequence of the double
counting procedure
employed in Ref.\ \cite{Koo11} and 
{\it not}
an artifact of the GGA
as stated there 
because our calculations
shown
in Fig.\  6 yield close values in the $\alpha$-region of
interest,
both for the LDA and the GGA. Also the RPA-derived value $\alpha^{\rm RPA}
 \sim 
1.4$ \cite{Enderle10}
could be approached for unrealistic
 small $U$-values below 
3~eV adopting the XRD data, only.

\section{\bf 4.\ SUMMARY}
The main result of our
revisited analysis
of \Livcu \ 
is the clear evidence
for 
 {\it strong}
 coupling of AHC
as derived from four independent
experimental and theoretical studies:
the INS  yields a dynamical
asymmetry parameter $\rho$
and a pitch angle very sensitive to quantum fluctuations.
Weak coupling would result in a nearly collinear 
incommensurate
state and in almost vanishing dynamical
anisotropy, i.e.\ 
$\rho \rightarrow 1$ not compatible with
 the 
diffraction and INS data.The obtained values for the
main exchange integrals are
supported by independent
microscopic
calculations
 based on the L(S)DA+$U$ approach and the 
multiband Hubbard model.


\acknowledgments
\noindent
We thank the
DFG 
[grant 
DR269/3-1 (S.-L.D. \& J.M.) and
the
Emmy-Noether-program (H.R.), 
the programs
PICS
[contracts 
CNRS 4767, NASU 243 (R.O.K)], and ASCR(AVOZ10100520) (J.M.)
for financial support.






\bibliography{99}

\end{document}